\documentclass[%
 preprint,
 amsmath,amssymb,
 aps,
 pre,
showkeys
]{revtex4-2}

\usepackage{graphicx}
\usepackage{dcolumn}
\usepackage{bm}
\usepackage{hyperref}
\usepackage{xcolor}
\usepackage{ctable}

\begin{document}

\title{Effect of nonzero temperature to condensed fraction of a homogeneous dilute weakly interacting Bose gas}

\author{Nguyen Van Thu and Pham Duy Thanh}
\affiliation{Department of Physics, Hanoi Pedagogical University 2, Hanoi 100000, Vietnam}
\email[]{nvthu@live.com}


\date{\today}

\begin{abstract}
We investigate the effect of non-zero temperature to the condensate fraction of a homogeneous dilute weakly interacting Bose gas in very low-temperature region. Within inproved Hartree-Fock approximation, the Cornwall–Jackiw–Tomboulis effective action approach shows that the thermal fluctuations make the condensate fraction decrease as a second and fourth-order power law of temperature. The result is compared with experimental data. Indeed, the effect of non-infinite size of the trap is also integrated.
\end{abstract}

\keywords{Quantum and thermal fluctuations, condensate fraction, Bose gas, improved Hartree-Fock approximation}

\maketitle

\section{Introduction\label{sec1}}

Bose-Einstein condensate (BEC) is a macroscopic coherent state of matter, which typically formed when the temperature of a system of bosonic atoms is cooled to bellow the critical temperature \cite{Bose,Einstein}. This state of matter was first predicted by Einstein in 1924. Seventy years later, the first BEC was created in experiment by Anderson {\it et al.} \cite{Anderson1995}. In this experiment, around 2000 atoms of isotope rubidium 87 was cooled by using magneto-optical traps along with laser and evaporative cooling techniques to the nanokelvin temperature scales required for the BEC. Anderson {\it et al.} found that below the critical temperature $T_C\approx 170$ nK, the BEC can be seen. Since this event, the number of studies of BEC have rapidly increased in both theory and experiment. Theoretically, the investigations on BEC at zero temperature have been attracted many attentions of physicists since Gross-Pitaevskii (GP) theory discovered \cite{Gross,Pitaevskii1}. The studies of BEC at finite temperature is also very rich (see \cite{Griffin2009} for review). Experimental works on BEC have also blossomed with creating single BEC and two-component BECs, particularly for BEC(s) of alkali atoms, such as Cs and Yb with both attractive (Cs + $^{174}$Yb) and  repulsive (Cs + $^{170}$Yb) interspecies interactions \cite{Wilson2021}, $^{41}$K and $^{87}$Rb in a hybrid trap \cite{Burchianti2018}, $^{39}$K and $^{87}$Rb \cite{Lee2018}, $^{23}$Na and $^{87}$Rb \cite{Wang2015}. The studies of BEC(s) provide an important tool for many aspects of physics, for example, quantum simulation \cite{Bloch2012} and sensing \cite{Cronin2009,Bongs2019}.

Theoretically, at zero temperature, all of bosonic atoms will be in the ground state. However, as we know, zero temperature cannot never be reached. Moreover, if we could reached zero temperature, a number of atoms are expelled from condensate due to quantum fluctuation \cite{Pethick}. This is surprising when we know about connection between superfluidity and BEC, which was posited by London \cite{London1938} and Tisza \cite{Tisza1947}. According to \cite{Miller1962,Glyde2011}, while at zero temperature liquid helium 4 is 100\% superfluid, less than 10\% of bosonic atoms are actually in BEC. To explain this fact, Bogoliubov proposed a microscopic theory for a homogeneous dilute weakly interacting Bose gas. This diluteness condition is characterized by the gas parameter, which is expressed via the particle density $n_0$ and the $s$-wave scattering length $a_s$, that is $\alpha_s=n_0a_s^3$. By this way, the gas can be seen as the diluteness if $\alpha_s\ll1$. The Bogoliubov's result for the condensed fraction at zero temperature is \cite{Bogolyubov},
\begin{eqnarray}
\frac{n_{\rm BEC}}{n_0}=1-\gamma_{\rm B}\alpha_s^{1/2},\label{Bogo}
\end{eqnarray}
in which $\gamma_{\rm B}=8/(3\sqrt{\pi})$. This result was also recovered authors latter, for example, within GP theory \cite{Stringari}, effective theory called the “simplified
approach” \cite{Carlen}, model with hard-sphere interaction between atoms \cite{Lee1} and so on \cite{Hohenberg,Horng}. Still at zero temperature, taking into account the higher-order of the gas parameter, the condensed fraction of the dilute weakly interacting BEC was calculated in form of (\ref{Bogo}) with a third term in right-hand side $-128\alpha_s/(3\pi)$ \cite{VanThu2022} within CJT effective action approach, $-32\alpha_s/(3\pi)$ \cite{Wu} by using the model where three particles interact
pairwise through the hard-sphere potential.

The condensed fraction (\ref{Bogo}) is pure theoretical. In reality, a BEC cloud produced in experiment is not homogeneous because of finite-size of the trap as well as we cannot reach zero temperature. Taking into account thermal fluctuations, Kehr \cite{Kehr1969} used the structure of the low-lying single-particle excitations of a Bose-condensed system to derive the $T^2$ contribution of temperature on the thermal depletion for Bose fluid at low temperatures. Applying this result for the dilute weakly interacting Bose gas, the result is
\begin{eqnarray}
\frac{n_{\rm BEC}}{n_0}=1-\gamma_B\alpha_s^{1/2}-\frac{m}{12\hbar^3cn_0}(k_BT)^2,\label{Shi}
\end{eqnarray}
with $\hbar, k_B$ being Boltzmann and reduced Planck constants, $m$ and $c$ atomic mass and speed of sound in the Bose gas, respectively. This correction has also reattained by Shi {\it et al.} \cite{Shi2000} within Popov approximation. In Bogoliubov approximation, a correction term for the thermal fluctuations is proportional to $T^4$ in limit of low-temperature \cite{Andersen}. In our previous work \cite{Thu2023a}, the non-condensate fraction due to thermal excitations was found in high-temperature limit. Our result shows the thermal contribution in two terms, which are proportional to the half-integer power law of temperature
\begin{eqnarray}
\frac{n_{\rm BEC}}{n_0}\approx1-\gamma_B\alpha_s^{1/2}-\frac{m^{3/2}\zeta(3/2)}{2\sqrt{2}\pi^{3/2}\hbar^3n_0}(k_BT)^{3/2}
+\frac{\sqrt{2}m^{3/2}\zeta(3/2)}{\pi^2\hbar^3n_0}(k_BT)^{3/2}\alpha_s^{1/2}.\label{nex}
\end{eqnarray}

The effect of non-infinite size in experiment was evaluated recently \cite{Thu2023} by means of Cornwall–Jackiw–Tomboulis (CJT) effective action approach within improved Hartree-Fock (IHF) approximation. It shows that the condensed fraction is decreased and this contribution can be neglected when the size $\ell$ of trap is much bigger than the healing length $\xi$ of Bose gas
\begin{eqnarray}
\frac{n_{\rm BEC}}{n_0}\approx1-\gamma_B\alpha_s^{1/2}-\frac{2\sqrt{\pi}}{3(\ell/\xi)}\alpha_s^{1/2}.\label{size}
\end{eqnarray}

The main aim of the present paper is to investigate the condensed fraction of a dilute weakly interacting Bose gas within IHF approximation at low twmpwrature.

This paper is organised as follows. In Section \ref{sec:2} we calculate the gap and Schwinger-Dyson (SD) equations for a single Bose gas in the IHF approximation by first recapitulating the regular HF method and then calculating these expressions for the CJT effective potential with symmetry-restoring terms. The condensate fraction at low temperature is investigated in Section \ref{sec:3} and the results are compared with both related approaches and experimental data. Finally, we present the conclusions and a future outlook in Section \ref{sec:4}.

\section{The equations of state in the IHF approximation}\label{sec:2}

In this Section, we will establish the equations of state for a Bose gas, which are the gap and SD equations. To do so, we first derive the CJT effective potential \cite{CJT} in the IHF approximation.

We set the stage for our calculations by starting with a dilute Bose gas described by the following Lagrangian density \cite{Pethick},
\begin{equation}
{\cal L}=\psi^*\left(-i\hbar\frac{\partial}{\partial t}-\frac{\hbar^2}{2m}\nabla^2\right)\psi-\mu\left|\psi\right|^2+\frac{g}{2}\left|\psi\right|^4,\label{eq:1}
\end{equation}
wherein $\mu$ is the chemical potential. The field operator $\psi(\vec{r},t)$ depends on both the coordinate $\vec{r}$ and time $t$. According to Ref. \cite{Pethick}, the interaction potential between the atoms can be chosen as the hard-sphere model. In this case, the strength of the interaction between the atoms is determined by the coupling constant $g=4\pi\hbar^2a_s/m$, which is expressed in terms of the $s$-wave scattering length $a_s$ by making use of the Born approximation. Now, thermodynamic stability requires that $g>0$, i.e., the boson interactions are repulsive.

Let $\psi_0$ be the expectation value of the field operator in the tree-approximation, the GP potential is then taken from \eqref{eq:1}
\begin{equation}
V_{GP}=-\mu\psi_0^2+\frac{g}{2}\psi_0^4.\label{eq:VGP}
\end{equation}
Note that henceforth the system is considered without any external fields. Furthermore, the non-macroscopic part of the condensate moves as a whole so that the lowest energy solution $\psi_0$ is real and plays the role of the order parameter. The square of the order parameter $\psi_0^2$ is the density of the condensate. Minimizing the potential \eqref{eq:VGP} with respect to the order parameter, one arrives at the gap equation
\begin{equation}
\psi_0(-\mu+g\psi_0^2)=0,\label{eq:gaptree}
\end{equation}
and hence, for the broken phase
\begin{equation}
\psi_0^2=\frac{\mu}{g}.\label{eq:psi0}
\end{equation}
In order to proceed within the framework of the HF approximation, the complex field operator $\psi$ should first be decomposed in terms of the order parameter $\psi_0$ and two real fields $\psi_1$ and $\psi_2$, which are associated with quantum fluctuations of the field \cite{Andersen}, i.e.,
\begin{equation}
\psi\rightarrow \psi_0+\frac{1}{\sqrt{2}}(\psi_1+i\psi_2).\label{eq:shift}
\end{equation}
Plugging equation \eqref{eq:shift} into the Lagrangian density \eqref{eq:1}, the interaction Lagrangian density, which describes the interaction between the real fields, in the HF approximation is obtained
\begin{equation}
{\cal L}_{int}=\frac{g}{2}\psi_0\psi_1(\psi_1^2+\psi_2^2)+\frac{g}{8}(\psi_1^2+\psi_2^2)^2.\label{eq:Lint}
\end{equation}
In the tree approximation one finds the gap equation \eqref{eq:psi0} and the propagator or Green's function
\begin{equation}
D_0(k)=\frac{1}{\omega_n^2+E_{\text{(tree)}}^2(k)}\left(
              \begin{array}{cc}
                \frac{\hbar^2k^2}{2m}-\mu+g\psi_0^2 & \omega_n \\
                -\omega_n &  \frac{\hbar^2k^2}{2m}-\mu+3g\psi_0^2\\
              \end{array}
            \right),\label{eq:protree}
\end{equation}
where $\omega_n$ is the $n$th Matsubara frequency for bosons. The latter is defined as $\omega_n=2\pi n/\beta$ where $n\in{\mathbb{Z}}$ with $k_B$ being the Boltzmann constant. The $E_{\text{(tree)}}(k)$ is the energy spectrum of the elementary excitation, which is determined by examining poles of the Green's function \cite{Andersen,Negele}. The result is
\begin{equation}
E_{\text{(tree)}}(k)=\sqrt{\frac{\hbar^2k^2}{2m}\left(\frac{\hbar^2k^2}{2m}+2g\psi_0^2\right)}.\label{dispertree}
\end{equation}
The spectrum Eq. (\ref{dispertree}) was first achieved by Bogoliubov \cite{Bogolyubov}. For small momenta, this equation is gapless and linear in $k$ and indicates the spontaneous $U(1)$ continuous symmetry breaking, creating a Nambu-Goldstone boson. To continue our discussion, we introduce the CJT effective potential in the HF approximation that can be constructed from the interaction Lagrangian density \eqref{eq:Lint} in the manner that was pointed out in \cite{Thu,Phat},
\begin{equation}
\begin{split}
V_\beta^{\text{(CJT)}} =&-\mu\psi_0^2 +\frac{g}{2}\psi_0^4+\frac{1}{2}\int_\beta \mbox{tr}\left[\ln G^{-1}(k)+D_0^{-1}(k)G(k)-{1\!\!1}\right]\\
&+\frac{3g}{8}(P_{11}^2+P_{22}^2)+\frac{g}{4}P_{11}P_{22}\, ,\label{eq:VHF}
\end{split}
\end{equation}
for which the functions $P_{11}$ and $P_{22}$ are
\begin{subequations}
    \begin{equation}
        \label{eq:P11}
        P_{11}=\int_\beta G_{11}(k)
    \end{equation}
    \begin{equation}
        \label{eq:P22}
        P_{22}=\int_\beta G_{22}(k)
    \end{equation}\label{P}
\end{subequations}
The Matsubara integrals in these expressions are defined as follows
\begin{equation}
    \label{eq:fk}
    \int_\beta f(k)=\frac{1}{\beta}\sum_{n=-\infty}^{+\infty}\int\frac{d^3\vec{k}}{(2\pi)^3}f(\omega_n,\vec{k})\, .
\end{equation}
Here $G(k)$ is the propagator or Green's function in the HF approximation, which can be obtained by minimizing the CJT effective potential \eqref{eq:VHF} with respect to the elements of the propagator. Performing these calculations results in the following expression for the inverse propagator
\begin{equation}
G^{-1}(k)=D_0^{-1}(k)+\Pi,\label{eq:r11}
\end{equation}
in which
\begin{equation}
\Pi=\left(
              \begin{array}{cc}
                \Pi_1& 0 \\
                0 & \Pi_2\\
              \end{array}
            \right),\label{eq:r12}
\end{equation}
with the matrix entries $\Pi_1$ and $\Pi_1$ being the self-energies that can be constructed from \eqref{eq:P11} and \eqref{eq:P22}, i.e.,
\begin{subequations}
    \begin{equation}
        \label{eq:r13}
        \Pi_1=\frac{3g}{2}P_{11}+\frac{g}{2}P_{22}
    \end{equation}
    \begin{equation}
        \label{eq:r14}
        \Pi_2=\frac{g}{2}P_{11}+\frac{3g}{2}P_{22}\, .
    \end{equation}
\end{subequations}
The gap equation in the HF approximation can now be found by minimizing the CJT effective potential \eqref{eq:VHF} with respect to the order parameter $\psi_0$, i.e.,
\begin{equation}
-\mu+g\psi_0^2+\Pi_1=0.\label{eq:r15}
\end{equation}
Combining equations \eqref{eq:r11}-\eqref{eq:r15}, one has the propagator in the HF approximation
\begin{equation}
\label{eq:inverse_G_CJT}
G(k)=\frac{1}{\omega_n^2+E_{\text{(HF)}}^2(k)}\left(
              \begin{array}{cc}
                \frac{\hbar^2k^2}{2m}-\mu+g\psi_0^2+\Pi_2 & -\omega_n \\
                \omega_n &  \frac{\hbar^2k^2}{2m}-\mu+3g\psi_0^2+\Pi_1\\
              \end{array}
            \right),
\end{equation}
and consequently the dispersion relation in this approximation is
    \begin{equation}
        E_{\text{(HF)}}(k)=\sqrt{\left(\frac{\hbar^2k^2}{2m}-\mu+3g\psi_0^2+\Pi_1\right)\left(\frac{\hbar^2k^2}{2m}-\mu+g\psi_0^2+\Pi_2\right)}.\label{eq:r16}
    \end{equation}
The Goldstone theorem \cite{Goldstone} requires a gapless excitation spectrum. In the physics of BEC, this theorem is referred to as the Hugenholtz-Pines theorem \cite{Hugenholtz} at zero temperature and a more general proof for all value of temperature was given by Hohenberg  and Martin \cite{Hohenberg}. More discussion about the gapless excitation spectrum for the BEC at finite temperature were mentioned in Refs. \cite{Hutchinson,Griffin}. Equations \eqref{eq:r15} and \eqref{eq:r16} show a non-gapless spectrum, which means that the Goldstone theorem is violated in the case of spontaneously broken symmetry within the HF approximation. To restore the Nambu-Goldstone boson, we now employ the method developed in \cite{Ivanov}. In this way, a phenomenological symmetry-restoring  correction term $\Delta V$ need to be added into CJT effective potential, which has to contemporaneously satisfy three conditions:

(i) it restores the Goldstone boson in the condensed phase;

(ii) it does not change the HF equations for the mean field and;

(iii) the results in the phase of restored symmetry are not changed.\\
It is not difficult to check that this term has from
\begin{equation}
\Delta V=-\frac{g}{4}(P_{11}^2+P_{22}^2)+\frac{g}{2}P_{11}P_{22},\label{extra}
\end{equation}
Let the propagator in the IHF approximation, be denoted as $D_{\text{(IHF)}}(k)$, the CJT effective potential \eqref{eq:VHF} now becomes
\begin{eqnarray}
    \label{eq:VIHF}
        \widetilde{V}_\beta^{\text{(CJT)}}& =&  -\mu\psi_0^2 +\frac{g}{2}\psi_0^4+\frac{1}{2}\int_\beta \mbox{tr}\left[\ln D^{-1}_{\text{(IHF)}}(k)+D_0^{-1}(k)D_{\text{(IHF)}}(k)-{1\!\!1}\right]\nonumber\\
&&+\frac{g}{8}(P_{11}^2+P_{22}^2)+\frac{3g}{4}P_{11}P_{22}.
\end{eqnarray}
From the CJT effective potential in the IHF approximation \eqref{eq:VIHF}, one arrives at the gap equation
\begin{equation}
-\mu+g\psi_0^2+\Sigma_1=0,\label{eq:gap}
\end{equation}
and the SD equation
\begin{equation}
M=-\mu+3g\psi_0^2+\Sigma_2,\label{eq:SD}
\end{equation}
in which the self-energies $\Sigma_1$ and $\Sigma_2$ are
\begin{subequations}
    \begin{equation}
        \label{eq:sig1}
        \Sigma_1=\frac{3g}{2}P_{11}+\frac{g}{2}P_{22}
    \end{equation}
    \begin{equation}
        \label{eq:sig2}
        \Sigma_2=\frac{g}{2}P_{11}+\frac{3g}{2}P_{22}\, .
    \end{equation}
\end{subequations}

Combining equations \eqref{eq:VIHF}-\eqref{eq:sig2}, one can once again calculate the propagator, now in the IHF approximation, i.e.,
\begin{equation}
D_{\text{(IHF)}}(k)=\frac{1}{\omega_n^2+E_{\text{(IHF)}}^2(k)}\left(
              \begin{array}{lr}
                \frac{\hbar^2k^2}{2m} & \omega_n \\
                -\omega_n & \frac{\hbar^2k^2}{2m}+M \\
              \end{array}
            \right).\label{eq:proIHF}
\end{equation}
Hence, the resulting dispersion relation is
\begin{equation}
E_{\rm{(IHF)}}(k)=\sqrt{\frac{\hbar^2k^2}{2m}\left(\frac{\hbar^2k^2}{2m}+M\right)}.\label{disperIHF}
\end{equation}
Clearly, the Goldstone boson is restored in this approximation. This is precisely the reason why this approximation is called the \textsl{improved} Hartree-Fock approximation. The gap and SD equations \eqref{eq:gap} and \eqref{eq:SD}, together with the momentum integrals \eqref{tichphan} form the equations of state, which govern the variation of all quantities of the system.

To do further, we first deal with the momentum integrals in the gap and SD equations. Without loss of generality, henceforth we use the abbreviation $E(k)$ replacing for $E_{\rm (IHF)}(k)$.   In the IHF approximation, the momentum integrals are obtained from equations \eqref{P} after replacing $G(k)$ by $D(k)$. Using the following formula \cite{Schmitt},
\begin{equation}
\sum_{n=-\infty}^{+\infty}\frac{1}{\omega_n^2+E^2(k)}=\frac{\beta}{2E(k)}\left[1+\frac{2}{e^{\beta E(k)}-1}\right],
\end{equation}
one has
\begin{subequations}
\begin{eqnarray}
P_{11}&=&\int\frac{d^3\vec{k}}{(2\pi)^3}\frac{1}{2E(k)}\frac{\hbar^2k^2}{2m}+\int\frac{d^3\vec{k}}{(2\pi)^3}\frac{1}{E(k)}\frac{1}{e^{\beta E(k)}-1}\frac{\hbar^2k^2}{2m},\label{tichphan1}\\
P_{22}&=&\int\frac{d^3\vec{k}}{(2\pi)^3}\frac{1}{2E(k)}\left(\frac{\hbar^2k^2}{2m}+M\right)+\int\frac{d^3\vec{k}}{(2\pi)^3}\frac{1}{E(k)}\frac{1}{e^{\beta E(k)}-1}\left(\frac{\hbar^2k^2}{2m}+M\right).\label{tichphan2}
\end{eqnarray}\label{tichphan}
\end{subequations}
The first terms in right-hand side of (\ref{tichphan1}) and (\ref{tichphan2}) do not explicitly depend on the temperature. These integrals are ultraviolet divergent. This divergence are avoidable by means of the dimensional regularization \cite{Andersen} and thus, the integrals can be computed. The integral $I_{m,n}$ is
\begin{eqnarray}
        I_{m,n}({\cal M})&=&\int\frac{d^d\kappa}{(2\pi)^d}\frac{\kappa^{2m-n}}{(\kappa^2+{\cal M}^2)^{n/2}}\nonumber\\
        &=&\frac{\Omega_d}{(2\pi)^d}\Lambda^{2\epsilon}{\cal M}^{d+2(m-n)}\frac{\Gamma\left(\frac{d-n}{2}+m\right)\Gamma\left(n-m-\frac{d}{2}\right)}{2\Gamma\left(\frac{n}{2}\right)},\label{eq:tp}
\end{eqnarray}
where $\Gamma(x)$ is the gamma function, $\Omega_d=2\pi^{d/2}/\Gamma(d/2)$ is the surface area of a $d-$dimensional sphere and $\Lambda$ is a renormalization scale that ensures the integral has the correct canonical dimension. This number is usually absorbed into the measure, hence it will not appear in the results.  Applying \eqref{eq:tp} to \eqref{tichphan} with $d=3$, one finds
\begin{subequations}
    \begin{eqnarray}
    P_{110}&\equiv&\int\frac{d^3\vec{k}}{(2\pi)^3}\frac{1}{2E(k)}\frac{\hbar^2k^2}{2m}=\frac{(2m)^{3/2}M^{3/2}}{6\pi^2\hbar^3},\label{tichphan01}\\
    P_{220}&\equiv&\int\frac{d^3\vec{k}}{(2\pi)^3}\frac{1}{2E(k)}\left(\frac{\hbar^2k^2}{2m}+M\right)=-\frac{(2m)^{3/2}M^{3/2}}{12\pi^2\hbar^3}.\label{tichphan20}
    \end{eqnarray}\label{tichphan0}
\end{subequations}

Generally, the temperature-dependent integrals (\ref{tichphan}) cannot be calculated analytically. We can solve by either numerical method or in some limit cases.

We will now look for the quantum and thermal fluctuations and calculate the non-condensate fraction of the dilute Bose gas.

\section{The non-condensate fraction of a homogeneous dilute Bose gas\label{sec:3}}

Let us now investigate the non-condensate fraction of the dilute weakly interacting Bose gas in the IHF approximation. To this end, we first consider the pressure. The pressure is defined as the negative of the CJT effective potential \eqref{eq:VIHF} at the minimum, i.e. satisfying both the gap and SD equations
\begin{equation}
{\cal P}=-\widetilde{V}_\beta\bigg|_{\mbox{minimum}}\equiv -\widetilde{{\cal V}}_\beta^{\text{(CJT)}}.\label{eq:press}
\end{equation}
Now, substituting equations \eqref{eq:gap} and \eqref{eq:SD} into \eqref{eq:VIHF}, one has
\begin{eqnarray}
    \label{eq:V1}
        \widetilde{{\cal V}}_\beta^{\text{(CJT)}}&=&-\mu\psi_0^2+\frac{g}{2}\psi_0^4+\frac{1}{2}\int_\beta \mbox{tr}\ln D^{-1}_{\text{(IHF)}}(k)\nonumber\\
        &&+\frac{1}{2}(3g\psi_0^2-\mu-M^2)P_{11}\nonumber
    +\frac{1}{2}(g\psi_0^2-\mu)P_{22}^2\\
    &&+\frac{g}{8}(P_{11}^2+P_{22}^2)+\frac{3g}{4}P_{11}P_{22}.
\end{eqnarray}
The chemical potential is defined as first derivative of the pressure with respect to the condensate density \cite{Landau1980}
\begin{eqnarray}
\mu=\frac{\partial \cal P}{\partial n}.\label{chemical}
\end{eqnarray}
Plugging (\ref{eq:V1}) into (\ref{chemical}) one can find the chemical potential beyond the mean-field theory
\begin{equation}
\mu=gn_0+gP_{11}.\label{eq:chemical}
\end{equation}
Combining (\ref{eq:chemical}), (\ref{eq:gap}) and (\ref{eq:SD}) leads to the gap and SD equations in higher precision
\begin{eqnarray}
&&-1+\frac{n_{\rm BEC}}{n_0}+\frac{1}{2n_0}(P_{11}+P_{22})=0,\nonumber\\
&&{\cal M}=-1+3\frac{n_{\rm BEC}}{n_0}-\frac{1}{2n_0}(P_{11}-3P_{22}),\label{eqnew}
\end{eqnarray}
in which we defined ${\cal M}=M/gn_0$.

To process further, we now calculate the temperature-dependent parts of the momentum integrals (\ref{tichphan}). At low temperature, the thermal excitations has so small momenta that the integrals are convergent. The results are
\begin{eqnarray}
P_{11T}&=&\frac{(2m)^{3/2}\pi^2}{30\hbar^3M^{5/2}}(k_BT)^4,\nonumber\\
P_{22T}&=&\frac{(2m)^{3/2}}{12\hbar^3M^{1/2}}(k_BT)^2.\label{tichphanTT}
\end{eqnarray}
In sum, at low temperature, the momentum integrals (\ref{tichphan}) can be expressed
\begin{eqnarray}
P_{11}&=&\frac{(2m)^{3/2}M^{3/2}}{6\pi^2\hbar^3}+\frac{(2m)^{3/2}\pi^2}{30\hbar^3M^{5/2}}(k_BT)^4,\nonumber\\
P_{22}&=&-\frac{(2m)^{3/2}M^{3/2}}{12\pi^2\hbar^3}+\frac{(2m)^{3/2}}{12\hbar^3M^{1/2}}(k_BT)^2.\label{chottichphan}
\end{eqnarray}

Inserting (\ref{chottichphan}) into (\ref{eqnew}) one can write the gap and SD equations in dimensionless form
\begin{subequations}\label{eqnew1}
\begin{eqnarray}
&&-1+\frac{n_{\rm BEC}}{n_0}+\frac{2\sqrt{2}}{3\sqrt{\pi}}{\cal M}^{3/2}\alpha_s^{1/2}+\frac{m^2(k_BT)^2}{12\sqrt{2\pi}\hbar^4n_0^{4/3}\alpha_s^{1/6}{\cal M}^{1/2}}+\frac{m^4(k_BT)^4}{480\sqrt{2\pi}\hbar^8n_0^{8/3}\alpha_s^{5/6}{\cal M}^{5/2}}=0,\nonumber\\\label{eqnew1a}\\
&&{\cal M}=-1+3\frac{n_{\rm BEC}}{n_0}-\frac{10\sqrt{2}}{3\sqrt{\pi}}{\cal M}^{3/2}\alpha_s^{1/2}+\frac{m^2(k_BT)^2}{4\sqrt{2\pi}\hbar^4n_0^{4/3}\alpha_s^{1/6}{\cal M}^{1/2}}-\frac{m^4(k_BT)^4}{480\sqrt{2\pi}\hbar^8n_0^{8/3}\alpha_s^{5/6}{\cal M}^{5/2}}.\nonumber\\\label{eqnew1b}
\end{eqnarray}
\end{subequations}
It is clearly that (\ref{eqnew1}) is a set of two equations for two unknowns, in particular, the condensate fraction and effective mass.  Using iterative method, the condensed fraction can be expressed in term
\begin{eqnarray}
\frac{n_{\rm BEC}}{n_0}\approx 1-\gamma_{\rm B}\alpha_s^{1/2}-\frac{2m^2\alpha_s^{1/3}}{9\pi\hbar^4n_0^{4/3}}(k_BT)^2-\frac{m^4\alpha_s^{1/6}}{9\pi^{3/2}\hbar^8n_0^{8/3}}(k_BT)^4.\label{dens}
\end{eqnarray}
The condensed fraction (\ref{dens}) consists of two component: (i) the effect of pure quantum fluctuations is shown by second term in right-hand side, which coincides to the one by other works; (ii) the nonzero temperature results to two last terms in right-hand side and, of course, at zero temperature these terms vanish. Our result is also quite different from that obtained by Shi {\it et al.} (\ref{Shi}) if we note that the speed of sound in Bose gas depends on the gas parameter $c=\sqrt{gn_{\rm BEC}/m}$. In addition, our result contains both second and fourth-order terms of temperature.

It is very easy to see that the three-last terms in right-hand side of Eq. (\ref{dens}) express the non-condensate fraction due to the quantum-induced interaction and thermal excitations
\begin{eqnarray}
\frac{n_{\rm ex}}{n_0}=\gamma_B\alpha_s^{1/2}+\frac{2m^2\alpha_s^{1/3}}{9\pi\hbar^4n_0^{4/3}}(k_BT)^2+\frac{m^4\alpha_s^{1/6}}{9\pi^{3/2}\hbar^8n_0^{8/3}}(k_BT)^4.\label{nex}
\end{eqnarray}
\begin{figure}
  \includegraphics[width = 0.8\linewidth]{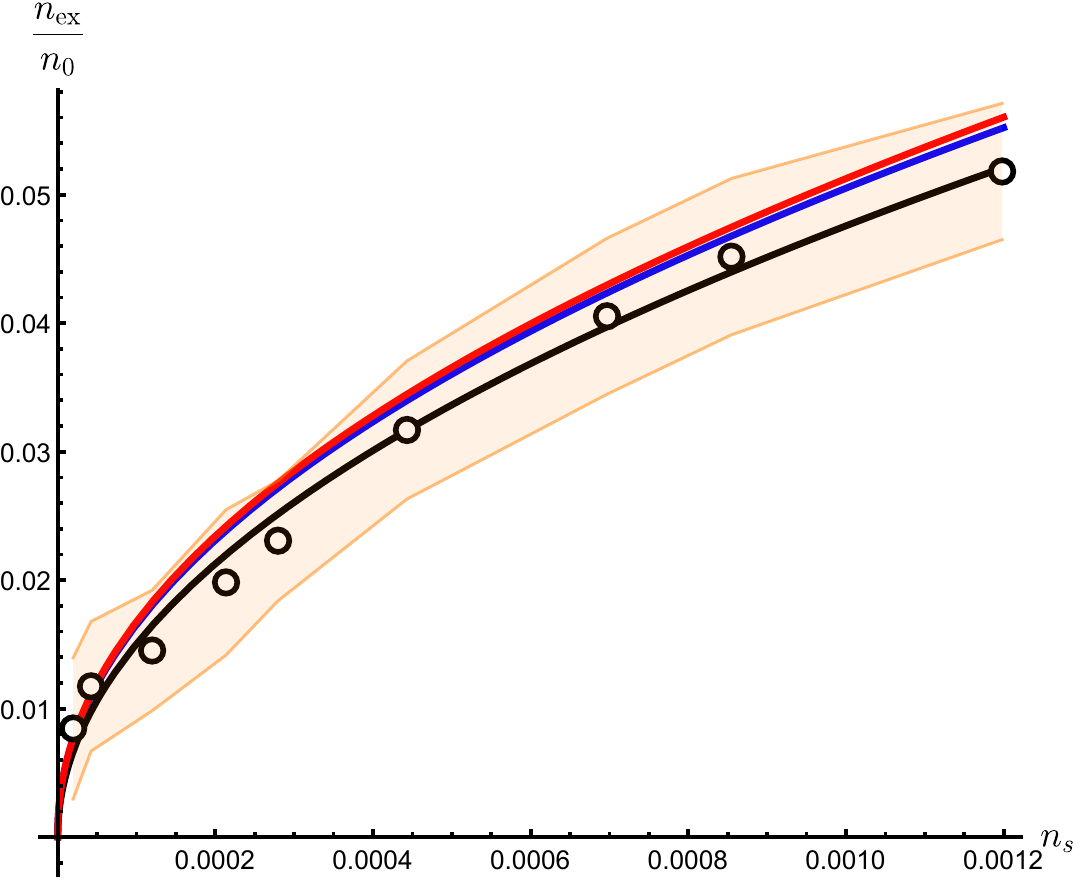}
\caption{(Color online) The non-condensate fraction $n_{ex}/n_0$ of a homogeneous Bose gas as a function of the gas parameter $\alpha_s$. The red line corresponds to Bogoliubov's result and blue curve expresses our result in Eq. (\ref{nex}) at temperature $T=3.5$ nK. The dots are experimental data \cite{Lopes} together with a orange error band.}
\label{fig:f1}       
\end{figure}
In order to compare with the experimental data we employ the data of Lopes {\it et al.} in Ref. \cite{Lopes}. In this experiment, isotope potassium 39 ($^{39}$K) was cooled in a cylindrical box trap. At initial state, Bose gas was produced at density of $n_0=3.5\times 10^{11}$ cm$^{-3}$ in the lowest hyperfine state $\left|F=1,m_F=1\right>$. As temperature was lowered to $T_C=20$ nK, a quasi-pure weakly interacting BEC clouds was formed at the scattering length $a_s=200a_0$, where $a_0=0.529$ \AA~ the Bohr radius, which satisfies the condition of diluteness $\alpha_s\ll1$. Atomic mass of the isotope $^{39}$K is of $m=38.963707u$ with $u=1.66053873\times 10^{-27}$ kg atomic mass unit. The non-condensate fraction was measured in range of time 150 - 250 ms during increasing of $s$-wave scattering length from $700a_0$ to $3000a_0$ at temperature in range from 3.5 nK to 5 nK.
The results are graphically shown in Fig. \ref{fig:f1}, in which the black curve corresponds to Bogoliubov's result (\ref{Bogo}) whereas the blue curve associates with (\ref{nex}) at $T=3.5$ nK. The experimental data are marked by open circles and error bar is the orange shading.

Now we consider the effect of both finite-size of the trap and non-zero temperature and assume that these effect are independent from each other. This assumption allows us to use   their additivity. Combining (\ref{nex}) and (\ref{size}) one arrives at
\begin{eqnarray}
\frac{n_{\rm ex}}{n_0}=\gamma_B\alpha_s^{1/2}+\frac{2m^2\alpha_s^{1/3}}{9\pi\hbar^4n_0^{4/3}}(k_BT)^2+\frac{m^4\alpha_s^{1/6}}{9\pi^{3/2}\hbar^8n_0^{8/3}}(k_BT)^4+\frac{2\sqrt{\pi}}{3(\ell/\xi)}\alpha_s^{1/2}.\label{nextotal}
\end{eqnarray}
The Eq. (\ref{nextotal}) is depicted by red curve in Fig. \ref{fig:f1} at finite-size of the trap is $\ell/\xi=50$. It is no doubt that not only thermal excitation but also finite-size of trap increase the non-condensate fraction.

\section{Conclusion and outlook\label{sec:4}}

In the forgoing Sections, the non-condensate fraction of the dilute weakly interacting Bose gas have investigated within IHF approximation by means of CJT effective action approach in low-temperature region. Our main results are in order

- The condensate fraction is contributed by two distinct kinds: quantum fluctuation on top of the ground state of the Bose gas. In case of the homogeneous gas, effect of non-zero temperature to condensate fraction can be expressed in terms of second and fourth-order of temperature.

- The effect of non-infinite size of trap is also calculated within assumption that it is independent of the temperature. Of course, it also makes the condensate fraction decrease.

It is interesting to study of both effects, in which these effects are not independent.

\begin{acknowledgements}
L. Raphael is acknowledged for providing and extensive discussions about the experimental data.
\end{acknowledgements}

\section*{Conflict of interest}
All of the authors declare that we have no conflict of interest.

\bibliography{bibliolow.bib}

\end{document}